# Status of the Spurious Evidence for Photoinduced Superconductivity

J. Steven Dodge, Leya Lopez, and Derek G. Sahota
Simon Fraser University, Burnaby, BC, V5T 2B2 Canada

*Abstract*—After more than a decade of research on photoinduced superconductivity, the experimental evidence for its existence remains controversial. Recently, we identified a fundamental flaw in the analysis of several influential results on $K_3C_{60}$ and showed that similar measurements on other compounds suffer from the same problem. We described how to account for this systematic error, and reanalyzed evidence that had previously been advanced for both photoinduced superconductivity and Higgs-mediated terahertz amplification. We found that both phenomena may be understood instead as a photoenhancement of the carrier mobility that saturates with fluence, with no need to appeal to a photoinduced phase transition to a superconducting state. We summarize this reinterpretation and describe how subsequent work on $K_3C_{60}$ provides quantitative support for it.

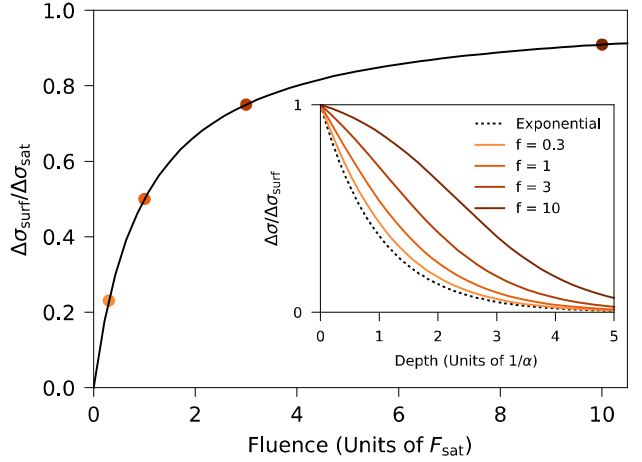

**Fig. 1.** Photoconductivity $\Delta\sigma$ as a function of fluence for a saturable medium. The main panel shows the surface photoconductivity $\Delta\sigma_{\text{surf}}$ normalized to its saturation value $\Delta\sigma_{\text{sat}}$. Markers indicate four values of the normalized fluence, $f = F/F_{\text{sat}}$, for which the inset shows the associated depth dependence of $\Delta\sigma$, normalized to the surface value (solid lines). The dotted line in the inset shows the exponential profile assumed in Refs. [3–7]. In a saturable medium, the effective thickness of the photoconducting layer increases logarithmically with fluence.

## I. INTRODUCTION

Recently we identified a large systematic error that corrupts the evidence for photoinduced superconductivity in $K_3C_{60}$ [1]. Most of this evidence has come from time-resolved terahertz (TR-THz) spectroscopy, which is sensitive to the nonequilibrium electrodynamic response of materials at the relevant frequencies and timescales [2–7]. However, TR-THz directly measures the complex reflection amplitude $r(\omega)$, not the local nonequilibrium complex conductivity $\sigma(\omega)$, and to relate them one needs to specify the complex photoconductivity depth profile $\Delta\sigma(\omega; z)$, which typically is not known independently. We showed that at the high excitation densities employed in these experiments, the photoconductivity depth profile must be distorted from the profile originally used to interpret the experiments. When we correct for this distortion, we obtain nonequilibrium conductivity spectra that are qualitatively different from those originally reported. The corrected results are consistent with a model in which photoexcitation enhances the carrier mobility but does not produce superconductivity. We expect this error to affect all the existing TR-THz evidence for photoinduced superconductivity, not just in $K_3C_{60}$. Below, we summarize the results of our previous analysis, discuss more recent results that provide additional support for it [7, 8], and respond to recent commentary on it [9].

## II. RESULTS

Previous work assumed that the refractive index of the photoexcited medium has the form [2–6],

$$n(\omega, z) = \bar{n}(\omega) + \Delta n_s(\omega) e^{-\alpha z}, \quad (1)$$

where $\bar{n}$ is the equilibrium refractive index, $\Delta n_s$ is the photoinduced change in the refractive index at the surface, $\omega$ is the probe frequency, $z$ is the depth from the surface, and $\alpha$ is the pump attenuation coefficient. This expression assumes that the energy density $\mathcal{E}$ absorbed by the pump decays as $\mathcal{E} \propto e^{-\alpha z}$, and that $n$ is linear in $\mathcal{E}$. But the photoresponse shows a *sublinear* dependence on the incident fluence $F$ in all the experiments that show evidence for photoinduced superconductivity [3–8], so analyzing these experiments in terms of Eq. (1) is not self-consistent.

To account for the observed nonlinearity, we consider the case of a local photoconductivity $\Delta\sigma$ that saturates with the absorbed energy density $\mathcal{E}$ while the pump absorption process remains linear, so that it retains the form $\mathcal{E} \propto e^{-\alpha z}$. Defining the dimensionless fluence parameter $f = F/F_{\text{sat}}$, where $F_{\text{sat}}$ is the characteristic scale for saturation, we express $\sigma$ as

$$\sigma(\omega, z, f) = \bar{\sigma}(\omega) + \Delta\sigma_{\text{sat}}(\omega) \frac{f e^{-\alpha z}}{1 + f e^{-\alpha z}}, \quad (2)$$

where $\bar{\sigma}$ denotes the equilibrium conductivity and $\Delta\sigma_{\text{sat}}$ denotes the saturated photoconductivity spectrum. Figure 1 shows the photoconductivity $\Delta\sigma$ as a function of $f$ and $z$ for fixed $\omega$, and reveals that the effective perturbation thickness increases with increasing $f$. We can define this thickness more precisely as $d_{\text{eff}} = \int dz\, \Delta\sigma(z)/\Delta\sigma_s$, where $\Delta\sigma_s$ denotes the complex photoconductivity at the surface, $z = 0$. For the profile in Eq. (1), $d_{\text{eff}} = \Lambda$, independent of fluence, but for the profile in Eq. (2), $d_{\text{eff}} = \Lambda \ln(1 + f)$. The pump penetration depth $\Lambda = 1/\alpha$ in $K_3C_{60}$ is about a third of the probe penetration depth, so the photoinduced change in the reflection amplitude is mainly sensitive to $\Delta G_\Box = \Delta\sigma_s d_{\text{eff}}$. Consequently, using Eq. (1) to deduce the surface conductivity for a medium that is really described by Eq. (2) will overestimate $\Delta\sigma_s$.

Figure 2 compares the photoexcited surface conductivity spectra originally reported by Budden et al. [5] and Buzzi et al. [6] with spectra that we have corrected using the observed nonlinearity [1]. For comparison, we also show a Drude-Lorentz fit to the equilibrium conductivity [6]. As expected, the corrected spectra exhibit smaller deviations from the equilibrium conductivity than the originally reported spectra. A Drude-Lorentz fit to the spectrum originally reported at $F = 3.0 \text{ mJ/cm}^2$ yields a Drude relaxation rate of $\gamma = 0$, as expected from a superconductor, but fits to the corrected spectra indicate a more moderate enhancement of the carrier mobility. We find that the Drude relaxation rate decreases by about a factor of 6 with increasing fluence, from $\hbar\gamma = 3.6$ meV in equilibrium to $\hbar\gamma = 0.6$ meV at $F = 4.5 \text{ mJ/cm}^2$, with $\hbar\gamma = 1.2$ meV at $F = 3.0 \text{ mJ/cm}^2$. Even in equilibrium, most of the spectral weight of the Drude peak is below the frequency range of the measurements, so this change in carrier mobility appears as a suppression of $\sigma_{s1}$ and an enhancement of $\sigma_{s2}$. Superficially, these changes resemble what we would expect if the photoexcited state were superconducting, but they will occur whenever photoexcitation causes the Drude peak to become narrower.

Our reanalysis also explains the negative real conductivity reported at $F = 4.5 \text{ mJ/cm}^2$, shown in Fig. 2(b), which was originally interpreted as evidence for Higgs-mediated optical parametric amplification. The original interpretation requires that the pump pulse produces a superconducting state at a temperature far above the equilibrium $T_C$, then the terahertz light field is amplified through a coupling to the Higgs field of this photoinduced superconductor. In our reinterpretation, this effect result from the systematic error in the analysis procedure, which conflates changes in the photoexcitation depth profile with changes in the surface conductivity.

More recent work has added support for our reanalysis. Rowe et al. [7] have reported measurements of $K_3C_{60}$ at $F = 18 \text{ mJ/cm}^2$ (see Fig. S6.2, $t = 0$) that are consistent with the trend shown in Fig. 2. We also obtain semiquantitative agreement with these results when we use our model to extrapolate the measurements of Ref. [5] at $F = 3.0 \text{ mJ/cm}^2$ to $F = 18 \text{ mJ/cm}^2$. Separately, Wang et al. [8] have reported results on $K_3C_{60}$ thin films that confirm our model for saturation. A fit to their results for the photoinduced impedance change, $\Delta Z$, as a function of fluence (Ref. [7], Fig. 5c) yields $F_{\text{sat}} = (1.08 \pm 0.15) \text{ mJ/cm}^2$, in quantitative agreement with the value $F_{\text{sat}} = (1.0 \pm 0.5) \text{ mJ/cm}^2$ that we obtained previously from other measurements [1].

Rowe et al. [7] continue to use Eq. (1) in their analysis but they also include another profile, in which $\Delta n \propto \sqrt{I}$, where $I$ is the pump intensity. Buzzi et al. [9] also rely on this profile in a recent commentary, where they note that it is motivated by assuming the conductivity is linear in the pump electric field. But as we discuss in Ref. [1] (Supplement, Sec. VI), existing evidence indicates that the photoconductivity is controlled primarily by the fluence, not the peak electric field, as this alternative profile assumes—moreover, it is unclear how the electric field, a vector quantity, would couple linearly to the scalar photoconductivity.

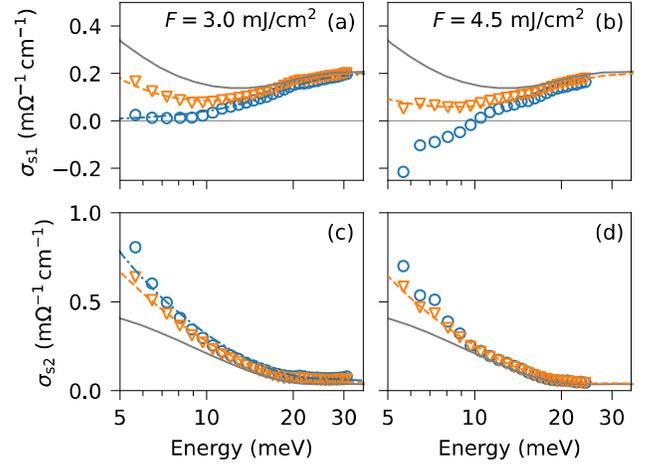

**Fig. 2.** Real (a,b) and imaginary (c,d) parts of the complex surface conductivity $\sigma_s$ for $F = 3.0 \text{ mJ/cm}^2$ (a,c) and $F = 4.5 \text{ mJ/cm}^2$ (b,d) with different profile assumptions. The spectra $\sigma_s(\omega; \mathcal{P}_{\text{exp}})$ reported by Budden et al. [5] and by Buzzi et al. [6] are shown as open circles in panels (a,c) and (b,d), respectively. The corrected spectra $\sigma_s(\omega; \mathcal{P}_{\text{sat}})$ are shown as open triangles. Lines show Drude-Lorentz fits to $\sigma_s(\omega; \mathcal{P}_{\text{sat}})$ (dashed) and $\bar{\sigma}(\omega)$ (solid). For the shaded region in (b) Buzzi et al. [6] reported negative real surface conductivity, which they interpreted as evidence for Higgs-mediated optical parametric amplification of terahertz radiation. This effect disappears when we account for the depth profile distortion that we derive from the observed sublinear fluence dependence.

## III. SUMMARY

Our reanalysis indicates that the experimental evidence for photoinduced superconductivity is distorted by a common systematic error. Future work is needed to critically review this literature and correct it for nonlinear distortion.